# TRISHUL: A Single-pass Optimal Two-level Inclusive Data Cache Hierarchy Selection Process for Real-time MPSoCs

Mohammad Shihabul Haque, Akash Kumar, Yajun Ha, Qiang Wu and Shaobo Luo
Department of Electrical and Computer Engineering, National University of Singapore
Email: {elemsh, akash, elehy, elewuqia, shaobo.luo}@nus.edu.sg

*Abstract*— Hitherto discovered approaches analyze the execution time of a real-time application on all the possible cache hierarchy setups to find the application specific optimal two-level inclusive data cache hierarchy to reduce cost, space and energy consumption while satisfying the time deadline in real-time Multi-Processor Systems on Chip (MPSoC). These brute-force like approaches can take years to complete. Alternatively, application's memory access trace driven crude estimation methods can find a cache hierarchy quickly by compromising the accuracy of results. In this article, for the first time, we propose a fast and accurate application's trace driven approach to find the optimal real-time application specific two-level inclusive data cache hierarchy. Our proposed approach "TRISHUL" predicts the optimal cache hierarchy performance first and then utilizes that information to find the optimal cache hierarchy quickly. TRISHUL can suggest a cache hierarchy, which has up to 128 times smaller size, up to 7 times faster compared to the suggestion of the state-of-the-art crude trace driven two-level inclusive cache hierarchy selection approach for the application traces analyzed.

## I. INTRODUCTION

Guaranteed execution time and performance in real-time computer applications allow planning the efficient use of the application as well as other related tasks. Due to this fact, from saving lives in hospitals to compressing images on digital cameras, real-time applications can be found everywhere. To satisfy the performance and time critical nature in the real-time applications, use of MPSoCs with multi-level cache hierarchy on real-time systems is growing day by day. By keeping data handy to the processors, cache memory hierarchy hides the latency of slow memory transactions. However, if the cache configurations[1] in the cache hierarchy are not chosen appropriately, it can have catastrophic effects by exceeding completion time deadline and by causing adverse effects on cost, space and energy consumption [2].

As a data cache hierarchy can have single or multiple cache memories in each level and one cache memory can influence others' cache hits/misses (inter-influencing), analyzing the given application's execution time on all possible cache hierarchy configurations[2] is a mandatory step in deciding the optimal application specific data cache hierarchy for real-time MPSoCs. However, take the cache hierarchy of Figure 1 (collected from [14]) to understand the problem with analysis time. Figure 1 depicts a widely used two-level inclusive data cache hierarchy (Harvard Architecture) on contemporary MPSoCs [8, 3, 11, 16]. In Figure 1, the processor cores include private caches which loads data in the shared cache before loading on them. The private caches search data in the shared cache before memory. Therefore, shared cache contains the superset of the private caches. See [22] for inclusive cache hierarchy details. If each cache memory has ten possible configurations and fifteen seconds are taken on average to find the execution time of an application on one cache hierarchy configuration, it will take eighteen days of continuous analysis to find the optimal cache hierarchy, unless any speedup mechanism is used.

Application's total execution time as well as time spent on instruction/data memory operation can be calculated quite accurately from the number of cache hits and misses [9]. Therefore, finding the number of cache hits and misses in each level of the data cache hierarchy will be enough to find the most appropriate application specific data cache hierarchy. If the range of cache hits and misses for each level in the data cache hierarchy (or required cache hierarchy performance *CHP*) to satisfy the allowable data memory operation time (*WCDMOT*) for real-time application is known, the searching process for the most appropriate cache hierarchy can be shortened by pruning the infeasible cache hierarchy configurations. Even though maximum allowable *WCDMOT* can be calculated using worst-case timing analysis [20, 17], to the best of our knowledge, no proposal has ever been made to estimate/predict the required performance of a multi-level/two-level inclusive data cache hierarchy in real-time MPSoCs. Moreover, no significantly fast method is known to find the most optimal two-level inclusive data cache hierarchy in real-time MPSoCs.

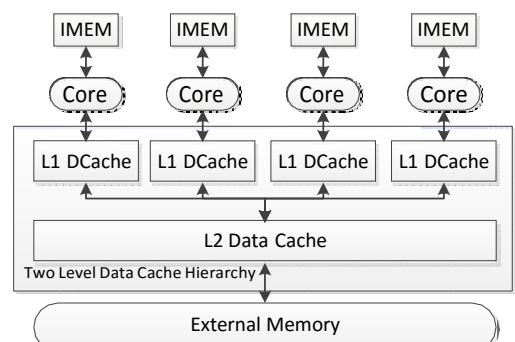

Fig. 1. Two-level Cache Hierarchy in MPSoC Architecture (collected from [14])

In this article, for the first time, we present a fast and accurate application's memory access trace driven process to select the *optimal two-level inclusive data cache hierarchy* for *real-time MPSoCs*. Our proposed cache hierarchy selection process "TRISHUL" (Time Restricted Interconnected Simulation of Hierarchical-cache Utility Library) finds the smallest storage capacity configurations, to save cost and space-energy consumption while meeting the time deadline, for each cache memory in the cache hierarchy. Our target architecture is the one presented in Figure 1. TRISHUL predicts the required

---

[1]Combination of cache parameters such as number of cache sets (set size), number of storage locations in each set (associativity), capacity of each storage location (cache line/block size), etc.

[2]A cache hierarchy setup with a specific configuration per cache memory.

*CHP* first with a novel approach. *CHP* is then used to reduce the cache hierarchy design space by pruning the infeasible cache hierarchy configurations. Therefore, a significant amount of time can be saved. To analyze each cache memory's behavior with minimal memory consumption and without effecting the accuracy of analysis, TRISHUL adopts "Single-pass technique (details in Section II), through a layered approach. Another unique feature of TRISHUL is, when a cache hierarchy is selected for an application but the *WCDMOT* has reduced, the optimal cache hierarchy can be found with minimal cache simulation. Due to all these features, TRISHUL can find the most optimal cache hierarchy in similar or less time than the state-of-the-art application trace driven crude method DIMSim [14] to select a two-level inclusive data cache hierarchy in real-time MPSoCs. TRISHUL is upto 7 times faster than DIMSim and the TRISHUL suggested shared caches can be up to 128 times less in size than DIMSim's suggestions for the application traces presented in this article. Note that TRISHUL can find the optimal one among all those cache hierarchies which have the same block size/cache line size in a particular level. The article is written assuming that all cache configurations can have a fixed block size only.

The rest of the paper is structured as follows: Section II discusses the related works, Section III explains TRISHUL's working policy and implementations, Section IV discusses the results and analyzes the TRISHUL suggested cache hierarchies' optimality and Section V concludes the paper.

## II. RELATED WORK

The worst case execution time of an application and the maximum number of main memory accesses estimated using worst-case timing analysis [4, 5] serve as the required *CHP* to select a single application specific cache memory. Even though real-time systems are usually application specific [18, 9] and the maximum number of main memory accesses acceptable for the *WCDMOT* can be estimated using the worst-case timing analysis, inter-influencing cache memories in the multi-level data cache hierarchy do not allow required *CHP* to be extracted from the number of memory accesses. No other methods are known either to predict the required *CHP* for real-time application specific two-level inclusive data cache hierarchy.

A single application specific cache memory is selected by evaluating the applications execution time on a large group of cache configurations. For this purpose, three types of application's memory access trace driven cache behavior simulation approaches are very popular due their speed compared to cycle accurate simulators or instruction set simulators. In the type called the compressed trace simulation, redundant information is pruned to compress the memory access trace [13, 19]. In the second type called the parallel simulation, cache configurations are simulated in parallel by using parallel hardware to reduce the overall simulation time [1, 15]. In contrast to parallel simulation, one processing unit is used as optimally as possible in the third type called single-pass simulation [12, 9]. In Single-pass simulation, several cache configurations are simulated by reading the application's memory access trace once. To mimic the hardware behavior as minimal as possible, cache configurations are represented by mainly four cache parameters: (i) set size ($S$), (ii) associativity ($A$), (iii) cache block/line size ($B$) and (iv) replacement policy. To simulate all the cache configurations quickly and accurately, several additional mechanisms (such as Inclusion properties [7], Intersection properties [6], etc.) are applied too in single-pass simulation. Single-pass simulation can be deployed with compressed trace simulation and/or parallel simulation.

Due to the advantages, attempts have been made to adopt single-pass simulation techniques to select appropriate multi-level cache hierarchy. Two proposals made by Wei Zang et al.[22, 23] are the latest in these attempts. However, Zang's approaches are limited to two-level Exclusive Cache hierarchy[3] only. Cache coherency is not considered in Zang's approaches; hence, not usable in MPSoCs causing coherency through data sharing. Zang's approaches are very restricted in terms of usability as they require the cache hierarchy to have first level cache with Least Recently Used (LRU) replacement policy and the second level cache with First-In-First-Out (FIFO) replacement policy ( Note that TRISHUL allows different replacement policies in shared and private caches).

To the best of our knowledge, only one approach DIMSim has adopted single-pass simulation so far to make a crude selection of two-level inclusive data cache hierarchy in real-time MPSoCs. To handle coherency, DIMSim finds a shared cache first that can satisfy the *WCDMOT* alone and, on top of that, a private level configuration to cover system overheads. As a result, the size of the shared cache is always much larger than required. Moreover, due to addition of private caches, traffic to shared cache is reduced causing a reduction in memory operation time further. As system overheads are dynamic and not predictable, adding a private cache per processor to handle unpredictable amount of system overhead is impractical and can cause excessively large private caches. Most importantly, shared caches cannot be found using DIMSim if the *WCDMOT* is not large enough to be satisfied by a single cache (Section IV details this problem with experiment results).

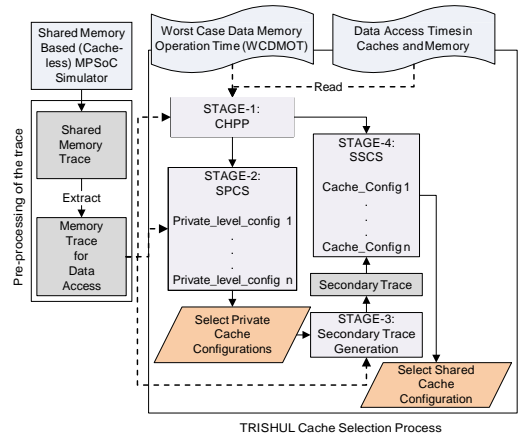

Fig. 2. TRISHUL Cache Hierarchy Selection Flow

## III. TRISHUL

In search for the optimal real-time application specific two-level inclusive data cache hierarchy, TRISHUL deployes three major components: (a) Cache Hierarchy Performance Predictor (CHPP), (b) Single-pass Private Cache Simulator (SPCS), and (c) Single-pass Shared Cache Simulator (SSCS). Figure 2 depicts the work flow of these components. The target real-time application's memory access trace is prepared beforehand. To generate the trace, memory accesses are observed and captured at the memory controller while the real-time MPSoC (without caches) is executing the application as communicating tasks or multiple applications. After that, data accesses from the trace are extracted and annotated; so that, for every

---

[3]In exclusive cache hierarchy, requested content is loaded in the privates caches directly from memory and shared cache stores the evicted content from private caches

322

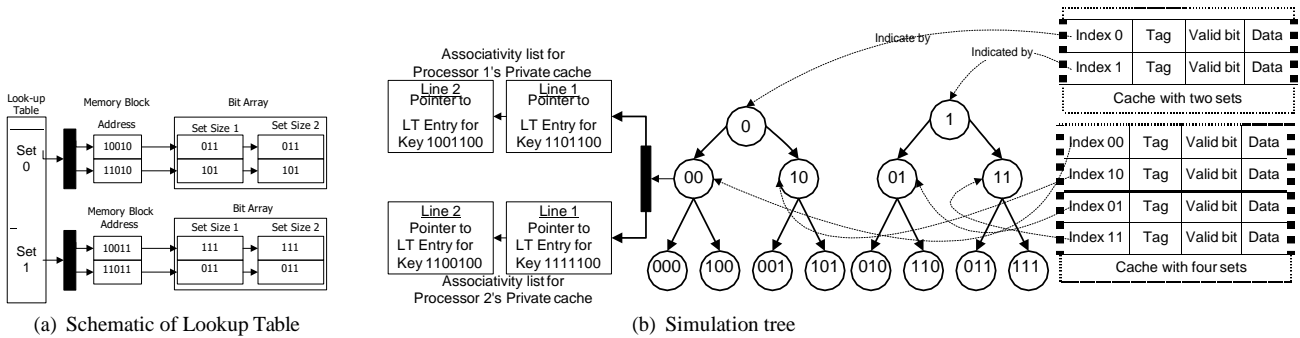

(a) Schematic of Lookup Table  (b) Simulation tree

Fig. 3. SPCS Data Structures

private cache. To reduce entry search time, look-up table entries are arranged into sets and entries are sorted on their keys to facilitate binary search. Memory blocks are mapped to look-up table sets just like cache sets.

By reading the trace entry for memory block ($RA$) once, SPCS determines cache misses in all the private level configurations by utilizing Algorithm 1. Just by reading the bit arrays for $RA$, SPCS identifies the appropriate processor's cache in each private level configuration that has not stored $RA$. For these caches, SPCS records misses and then updates the look-up table and simulation tree to reflect after miss scenario. To record the number of sequential data blocks served from the private level ($TAP^r$), SPCS just counts the number of different cycles when a data access occurred. From the bit arrays, SPCS also knows quickly which processors' caches have to be updated/invalidated when a particular processor updates a shared data (coherency handling).

After simulation, all the private level configurations' observed $TAP$ and $TAS$ are substituted in Equation 1 to find the largest value for $TAM$ below the CHPP predicted $TAM$ (we call this fine tuned value as $TAM^r$). The private level configuration that generates the largest $TAM^r$ is selected for shared cache generation and its $TAM^r$ is used as the miss limit in shared cache simulation.

A. Single-pass Shared Cache Simulator (SSCS)

*Role:* Finding the optimal shared cache configuration by reading the secondary trace file only once.

*Details:* SSCS simulates one shared cache memory's multiple configurations. SSCS is actually the simulator of [6] without any intersection property deployed and modified to accommodate the use of $TAM^r$. The Look-up table and simulation tree are also used by SSCS to represent shared cache configurations. However, one look-up table and its associated simulation tree are generated to simulate cache configurations with varying $S$ and $A$. The look-up table and the simulation tree in SSCS looks exactly same as in SPCS; however, the bits in the look-up table bit arrays provide data availability information for cache configurations with the same $S$ and $B$ but with different $A$. For example, when the look-up table in Figure 3(a) will be used for SSCS, the bit array associated with memory address 10010 for $S = 1$ will indicate that the memory block content will be absent in the shared cache configurations with $S = 1$ and $A = 1$ and 2 provided three options 1,2 and 4 for $A$ value. SSCS will add three lists containing 1, 2 and 4 nodes to represent $A = 1, 2$ and 4 respectively with each tree node. That means; the top level in the tree in Figure 3(b) will represent the fixed cache line sized shared cache configurations with $S = 2$ and $A = 1, 2$ and 4. After simulation, the shared cache configuration with the largest number of memory accesses ($TAM^{rr}$)

**Algorithm 1:** SPCSEvaluation(RequestedAddress(RA), RequestingProcessor(N), MissLimit(TAS))

1  $LT$ = Look-up Table;
2  $A_N$ = The associativity list for the $N^{th}$ processor;
3  **if** (RA is not found in LT) **then**
4     Record one cache miss for all the configurations of Processor $N$'s private cache;
5     Exclude the tree level $L$ from simulation whose total number of misses is greater than $TAS$
6     Place $RA$ in $LT$ and place pointer to $RA$'s location in $LT$ in all the
7     configurations for processor $N$'s private caches in the simulation tree;
8  **else**
9     Select the tree level $L = 0$ (smallest cache set size $S = 2^L$) in the tree;
10    **while** $2^L$ is not larger than the largest set size **do**
11       **if** (Lth level is not excluded for simulation) **then**
12         **if** (Write Operation) **then**
13           **if** (RA found in $A_N$) **then**
14             For set size $2^L$, update/invalidate bit arrays for $RA$ in $LT$ for
15             the processors $I$ where $I \neq N, I = 1, 2, 3, ..., $ Last processor;
16           **else**
17             Record a cache miss for processor $N$'s configuration with set size $2^L$;
18             Place $RA$ in $A_N$ and update the $LT$ record;
19             update/Invalidate bit arrays for $RA$ in $LT$ for
20             all $A_I$ where $I \neq N, I =$
             $1, 2, 3, ..., $ Last processor number;
21         **else**
22           **if** (Not found in $A_N$ or invalid) **then**
23             Record a cache miss for processor $N$'s configuration with set size $2^L$;
24             Place $RA$ in $A_N$ and update the $LT$ record;
25       Exclude the tree level $L$ from simulation whose total number of misses is greater than $TAS$
26     $L = L + 1$;

is selected for final design.

From here, we use $n$ apostrophes (') after $TAP$ and $TAS$ but $n + 1$ apostrophes after $TAM$ to indicate the observed number of sequential accesses in private level, shared level and main memory in the $n^{th}$ cache hierarchy configuration.

By now, readers may be starving to know, when TRISHUL selects a private level configuration $X$ with $TAS^r$ misses and a shared cache configuration with $TAM^{rr}$ misses:

1. Will there be any smaller private level configuration with $TAS^{rr} > TAS^r$ and larger shared cache configuration with $TAM^{rrr} \leq TAM^{rr}$ that still satisfy $WCDMOT$?

2. If no shared cache is found for $X$, can a larger private level configuration with $TAS^{rr} < TAS^r$ have a shared cache to satisfy $WCDMOT$?

3. How to select the optimal cache hierarchy with minimal simulation when an application's $WCDMOT$ reduces?

| Trace | WCDMOT (sec) | TRISHUL (Optimal) | | DIMSim | TRISHUL | | DIMSim |
|---|---|---|---|---|---|---|---|
| | | Private Config | Shared Config. | Shared Config | AMT (Sec) | Decision in (Sec) | Decision in (Sec) |
| JPEG | | | | | | | |
| barbara | 1.00 | (8X2) | (1X2) | (8X16) | 0.96 | 1700 | 1832 |
| | 0.40 | (4X16) | (1X2) | N/A | 0.40 | 361 | N/A |
| | 0.15 | (16X16) | (64X16) | N/A | 0.15 | 281 | N/A |
| criss | 1.00 | (8X2) | (1X2) | (8X16) | 0.96 | 1699 | 1758 |
| | 0.40 | (4X16) | (1X2) | N/A | 0.40 | 345 | N/A |
| | 0.15 | (16X16) | (64X16) | N/A | 0.15 | 280 | N/A |
| graph | 1.00 | (8X2) | (1X2) | (8X16) | 0.96 | 1792 | 1752 |
| | 0.40 | (4X16) | (1X2) | N/A | 0.40 | 354 | N/A |
| | 0.15 | (16X16) | (128X8) | N/A | 0.15 | 283 | N/A |
| lena | 1.00 | (8X2) | (1X2) | (8X16) | 0.96 | 1761 | 1735 |
| | 0.40 | (4X16) | (1X2) | N/A | 0.40 | 344 | N/A |
| | 0.15 | (16X16) | (64X16) | N/A | 0.15 | 281 | N/A |
| photo1 | 1.00 | (8X2) | (1X2) | (8X16) | 0.96 | 1769 | 1722 |
| | 0.40 | (4X16) | (1X2) | N/A | 0.40 | 346 | N/A |
| | 0.15 | (16X16) | (128X8) | N/A | 0.15 | 281 | N/A |
| photo2 | 1.00 | (8X2) | (1X2) | (8X16) | 0.96 | 1772 | 1751 |
| | 0.40 | (4X16) | (1X2) | N/A | 0.40 | 346 | N/A |
| | 0.15 | (16X16) | (128X8) | N/A | 0.15 | 281 | N/A |
| H264 | | | | | | | |
| Bluesky | 1.00 | (2X8) | (8X8) | (8X4) | 0.99 | 2336 | 1526 |
| | 0.75 | (8X4) | (1X2) | (16X16) | 0.61 | 525 | 1525 |
| | 0.40 | (2X16) | (64X16) | N/A | 0.39 | 511 | N/A |
| river | 1.00 | (2X8) | (8X8) | (8X4) | 0.99 | 2145 | 1541 |
| | 0.75 | (8X4) | (1X2) | (16X16) | 0.61 | 524 | 1472 |
| | 0.40 | (2X16) | (64X16) | N/A | 0.38 | 640 | N/A |
| station | 1.00 | (2X8) | (8X8) | (8X4) | 0.99 | 2255 | 1506 |
| | 0.75 | (8X4) | (1X2) | (16X16) | 0.61 | 600 | 1422 |
| | 0.40 | (2X16) | (64X16) | N/A | 0.38 | 478 | N/A |
| pedest. | 1.00 | (2X8) | (8X8) | (8X4) | 0.99 | 2262 | 1464 |
| | 0.75 | (8X4) | (1X2) | (16X16) | 0.61 | 529 | 1436 |
| | 0.40 | (2X16) | (64X16) | N/A | 0.38 | 696 | N/A |
| tractor | 1.00 | (2X8) | (8X8) | (8X4) | 0.99 | 2255 | 1507 |
| | 0.75 | (8X4) | (1X2) | (16X16) | 0.61 | 523 | 1449 |
| | 0.40 | (2X16) | (64X16) | N/A | 0.39 | 706 | N/A |

TABLE I
EFFICIENCY OF TRISHUL OVER DIMSIM

Let's answer all these questions in the following section.

## IV. EXPERIMENT AND RESULTS

DIMSim showed that a crude estimation of a real-time application specific two-level inclusive data cache hierarchy reduces the design space exploration time from years to minutes. Therefore, our experiment setup is to find out whether TRISHUL can find the optimal cache hierarchy within similar or less time than DIMSim. We implement TRISHUL using C language and re-implement DIMSim following the guidelines provided in [14].

We implement a six core cache-less multiprocessor implementation using the Tensilica tool set [21]. Like DIMSim, we execute JPEG encoder and H264 encoder (only the motion estimation kernel) to generate traces for different image and video benchmarks. Both the applications are partitioned into multiple communicating/sharing tasks which are mapped on separate processors. Data sharing is performed only through shared cache.

For Simulation, we execute TRISHUL and DIMSim on a machine with a dual core Opteron64 2GHz processor, 8GB of main memory and 1MByte shared L2 cache. In our experiment, each private cache or shared cache has 75 possible configurations where $S = 1$ to $16384$, $A = 1, 2, 4, 8, 16$, $B = 4Bytes$ and FIFO replacement policy. We used $TP = 1$ ns, $TS = 4$ ns, and $TM = 15$ ns (based on the Xtensa processor [21]), assuming that all the applications are mapped on a 1GHz processor with one clock cycle private cache latency.

Table I presents the experiment results in TRISHUL and DIMSim. Column 1 presents the six JPEG traces and five H264 traces. Column 2 presents the generous (1.0sec), regular (0.40sec for JPEG and 0.75sec for H264) and stingy (0.15sec for JPEG and 0.40sec for H264) $WCDMOT$ calculated using [10] for every trace file. Column 3 presents the configuration of each cache in the private level selected by TRISHUL. Column 4 and 5 present the shared cache configurations selected by TRISHUL and DIMSim respectively. No private level cache configuration has been presented for DIMSim as no practical private cache selection criteria has been provided in [14]. Column 6 presents the actual data operation time ($AMT$) of the TRISHUL selected cache hierarchy.

Column 7 presents the total time to select an optimal cache hierarchy in TRISHUL. The last column presents the time to select a shared cache only in DIMSim. For example, for JPEG Barbara and $WCDMOT = 1.0$sec, TRIHSUL selected a private level configuration with each cache containing ($S = 8$, $A = 2$ and $B = 4Bytes$). The selected shared cache configuration in TRISHUL and DIMSim contain ($S = 1$, $A = 2$ and $B = 4Bytes$) and ($S = 8$, $A = 16$ and $B = 4Bytes$) respectively. The TRISHUL selected cache hierarchy has a $AMT = 0.96$sec. For this case, TRISHUL selected the entire cache hierarchy in 28min (approx). On the other hand, DIMSim took almost 31min just to select a shared cache. Note that in this example the entire cache hierarchy selected by TRISHUL is only 396Bytes. However, DIMSim's shared cache is alone 512Bytes. The results reveal that TRISHUL selected shared cache can be 128 times smaller or 2 times bigger in size compared to DIMSim's shared cache. Readers may wandering why, sometimes DIMSim suggested shared cache is smaller than TRISHUL suggested shared cache in the cache hierarchy (ex. bluesky and $WCDMOT = 1.0$). In TRISHUL, private cache misses generated in parallel, are sequentialized and searched in shared cache. This ordering process is random. Depending on the ordering, cache misses may increase in the shared cache. In DIMSim, no parallel access is considered. Therefore, if the trace file has the most optimized ordering of accesses, DIMSim may produce smaller shared caches compared to TRISHUL. In Figure 4, the CHPP predicted values for $TAP$, $TAS$ and $TAM^r$ and their corresponding $TAP^r$, $TAS^r$ and $TAM^{rr}$ in the TRISHUL selected cache hierarchies are presented in groups for generous, regular and stingy $WCDMOT$. In each group, the order of the trace files is same as the order in Table I where the left most bar pair indicates the predicted and observed values in Barbara and the right most bar pair represents the tractor trace. Figure 4(b) shows that $TAS$ is within 96%-64% accuracy range compared to $TAS^r$. Similarly, Figure 4(c) shows that $TAM^r$ is within 99.95%-74.55% accuracy range compared to $TAM^{rr}$. The minimum value of sequential private cache accesses ($TAP$) is also very accurately predicted by CHPP (see in Figure 4(a)). Due to the accurate predictions, TRISHUL can select a cache hierarchy without simulating all the possible configurations for each cache. For each JPEG trace, TRISHUL simulated *neither more than 54 nor less than 38 out of 75 private level configurations*. For every H264 trace, the number of private level configurations simulated is in between *33 to 53*. TRISHUL simulated 27-60 and 30-45 shared cache configurations for JPEG and H264 respectively. Moreover, the cache hierarchy selected by TRISHUL for each trace file can closely satisfy the given $WCDMOT$ with their $AMT$.

Results show that TRISHUL is quite efficient in finding a cache hierarchy for the given criteria. However, to prove that TRISHUL choose the optimal cache hierarchy, we have to answer Question 1 and 2. To find the answers, let's take an example to analyze. Lets consider that we have two cache hierarchies 'H1' with private level configuration 'C1' and 'H2' with private level configuration 'C2'. 'C2' is bigger than 'C1'. For a trace file and fixed $B$, number of misses in 'C2' is $TAS^{rr}$ which will always be smaller than misses in 'C1' ($TAS^r$). For a fixed $B$ and trace file, total number of sequential accesses to the private level ($TAP$) does not change when cache hierarchy configuration is changed (see Figure 4(a)). Therefore,

1. *Answer for Question 1:* if both 'H1' and 'H2' could satisfy $WCDMOT$ with equal number of memory accesses ($TAM^{rr} = TAM^{rrr}$), it means $(TAP^r \times TP) +$

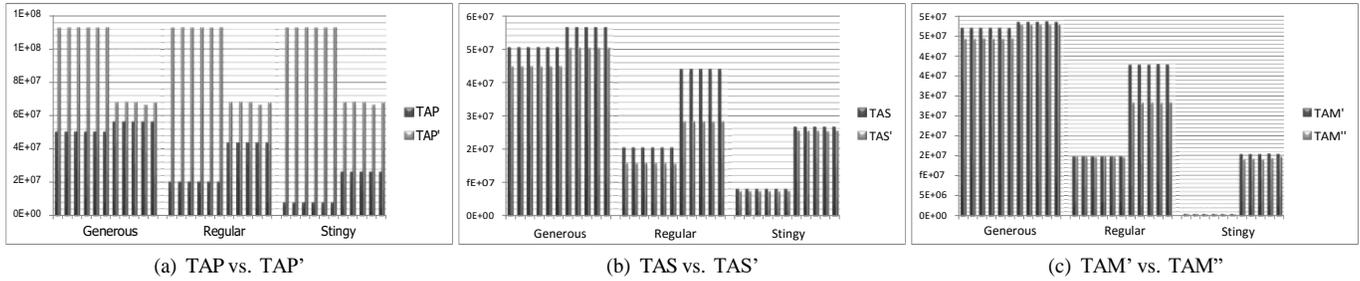

(a) TAP vs. TAP'    (b) TAS vs. TAS'    (c) TAM' vs. TAM''

Fig. 4. Predicted vs. Observed Values in TRISHUL

$(TAS^r \times TS) + (TAM^{rr} \times TM) = (TAP^{rr} \times TP) + (TAS^{rr} \times TS) + (TAM^{rrr} \times TM)$. As $TAP^r = TAP^{rr}$ and $TAM^{rr} = TAM^{rrr}$, $TAS^r$ must be equal to $TAS^{rr}$. Otherwise $WCDMOT$ cannot be satisfied. So, 'H1' cannot satisfy $WCDMOT$ when $TAS^r > TAS^{rr}$ and $TAM^{rr} = TAM^{rrr}$. Even if 'H1' could satisfy $WCDMOT$ with $TAS^r > TAS^{rr}$ and $TAM^{rr} < TAM^{rrr}$, it would not be optimal. Because, less memory accesses means more storage capacity in the cache hierarchy (more space and energy consumption besides being costly).

2. *Answer for Question 2:* In this case, TRISHUL selected private level configuration is representing 'C1' that cannot satisfy $WCDMOT$ with 'H1'. So, for 'H1', $WCDMOT - (TAP^r \times TP) < (TAS^r \times TS) + (TAM^{rr} \times TM)$. If 'H2' could satisfy $WCDMOT$, $WCDMOT - (TAP^{rr} \times TP) = (TAS^{rr} \times TS) + (TAM^{rrr} \times TM)$. As $TAP^r = TAP^{rr}$, it means $(TAS^{rr} \times TS) + (TAM^{rrr} \times TM) < (TAS^r \times TS) + (TAM^{rr} \times TM)$; or $TS \times (TAS^r - TAS^{rr}) > TM \times (TAM^{rrr} - TAM^{rr})$. As $TAS^r > TAS^{rr}$, to have a positive value of $(TAM^{rrr} - TAM^{rr})$, the $TAM^{rrr}$ must be larger than $TAM^{rr}$. But in reality, $TAM^{rrr}$ cannot be larger than $TAM^{rr}$ when $TAS^r > TAS^{rr}$. Because to satisfy $WCDMOT$ by 'H2', $TAM^{rrr}$ has to be less than $TAM^{rr}$. So, answer for Question 2 is "No".

So, TRISHUL selects the optimal cache hierarchy if there exists one.

From the last two columns of Table I, it can be seen that TRISHUL and DIMSim spent almost similar time to select a cache hierarchy and shared cache respectively for $WCDMOT = 1.0$sec. However, DIMSim failed to make any decision for $WCDMOT < 1.0$sec in any JPEG and for $WCDMOT < 0.75$sec in any H264 trace. The reason is, as DIMSim selects a shared cache first that alone can satisfy the given $WCDMOT$, it is impossible to satisfy $WCDMOT < 1.0$sec (for JPEG) or 0.75sec (for H264) by any single cache memory with any configuration simulated in our experiment. On the other hand, TRISHUL saves a huge amount of time for $WCDMOT < 1.0$sec. The reason is, when private level cache hierarchy configurations are simulated for $WCDMOT = 1.0$sec, SPCS records the results for any private level configuration that do not exceed the CHPP given $TAS$. Therefore, when the $WCDMOT$ reduces, required $TAS$ value will be decreased and can only be satisfied by a larger private level configuration. As all the larger private level configurations' $TAS^r$ values are recorded for the trace file in the SPCS produced result for $WCDMOT = 1.0$sec, the appropriate private level configuration can be selected without further simulation for any $WCDMOT < 1.0$sec with the help of CHPP. Once the private level configuration is selected, shared cache can be selected with the help of SSCS. This is the answer for Question 3. When DIMSim cannot find a shared cache for $WCDMOT < 1.0$sec in JPEG, the solution for $WCDMOT = 1.0$sec has to be used. Same goes for H264. For example, for $WCDMOT = 0.4$sec and bluesky, TRISHUL took around 8min to decide a cache hierarchy. But for DIMSim, the solution for $WCDMOT = 0.75$sec has to be used. So, DIMSim's decision time is 25min (3 times slower than TRISHUL). In this way, TRISHUL can be up to 7 times faster than DIMSIM for the traces analyzed in Table I.

V. CONCLUSION

In this article, we present an application trace driven method to select the optimal two-level inclusive data cache hierarchy selection process for real-time MPSoCs. The method TRISHUL presents a novel mechanism to find the required cache hierarchy performance without analyzing/simulating any cache memory behavior. TRISHUL can select an optimal cache hierarchy within a time period necessary to select a single shared cache by the available trace driven two-level inclusive data cache hierarchy selectors.